\documentclass[aps,prb,reprint,amsmath,superscriptaddress,amssymb,showpacs]{revtex4-1}
\usepackage[english]{babel}
\usepackage{graphicx}
\usepackage[utf8]{inputenc}
\usepackage{color}
\usepackage{hyperref}

\makeatletter
\renewcommand{\fnum@figure}{FIG. \thefigure}
\makeatother

\newcommand{\um}[1]{\,\mathrm{#1}}

\newcommand{\beq}{\begin{equation}}
\newcommand{\eeq}{\end{equation}}
\newcommand{\be}{\begin{equation}}
\newcommand{\ee}{\end{equation}}
\newcommand{\beqa}{\begin{eqnarray}}
\newcommand{\eeqa}{\end{eqnarray}}
\newcommand{\bea}{\begin{align}}
\newcommand{\eea}{\end{align}}

\newcommand{\vk}{\mathbf{k}}

\newcommand{\mb}[1]{\mathbf{#1}}
\newcommand{\dive}{\mathbf{\nabla}\cdot}
\newcommand{\grad}{\mathbf{\nabla}}
\newcommand{\rot}{\mathbf{\nabla}\times}
\newcommand{\St}[1]{\textrm{St}_{\textrm{#1}}}
\renewcommand{\rm}[1]{\textrm{#1}}

\begin{document}

\title{Polarity dependent heating at the phase interface in metal-insulator transitions}
\author{Giuliano Chiriacò}
\affiliation{Department of Physics, Columbia University, New York, NY 10027}
\author{Andrew J. Millis}
\affiliation{Department of Physics, Columbia University, New York, NY 10027}
\affiliation{Center for Computational Quantum Physics, The Flatiron Institute, New York, NY 10010}

\date{\today}

\begin{abstract}
Current-driven insulator-metal transitions are in many cases driven by Joule heating proportional to the square of the applied current. Recent nano-imaging experiments in Ca$_2$RuO$_4$ reveal a metal-insulator phase boundary that depends on the direction of an applied current, suggesting an important non-heating effect. Motivated by these results, we study the effects of an electric current in a system containing interfaces between metallic and insulating phases. Derivation of a heat balance equation from general macroscopic Onsager transport theory, reveals a heating term proportional to the product of the current across the interface and the discontinuity in the Seebeck coefficient, so that heat can either be generated or removed at an interface, depending on the direction of the current relative to the change in material properties. For parameters appropriate to Ca$_2$RuO$_4$, this heating can be comparable to or larger than Joule heating. A simplified model of the relevant experimental geometry is shown to provide results consistent with the experiments. Extension of the results to other inhomogeneous metal-insulator transition systems is discussed.
\end{abstract}

\maketitle

\section{Introduction}

Phase transitions induced by a non-equilibrium drive are a topic of fundamental importance and current experimental interest \cite{Aver:opt-rev,Basov:review,Mitra06,Aver:optCu,Cav:K3C60,laser:gold}. Transient and steady-state non-equilibrium drives may allow access to many phases, some of which are absent in equilibrium \cite{Kogar:CDW,FE:SrTiO,CMA:trsc,CMA:neg,Sun:neq,Mazur,VO2:science}. One important class of non-equilibrium transitions is the insulator-metal transition in a correlated electron insulator subject to a dc electric field or current \cite{Assamitsu97,Maecit:7,Maecit:8,Maecit:9,Maecit:10,Maecit:11}. The theoretical consensus is that the insulating phase may be destabilized when the driving field is such that the voltage drop over a unit cell is a significant fraction (greater than a few percent) of the insulating gap \cite{Amaricci12}. At these drives enough valence band carriers are excited over the gap to destroy the insulator, by either Landau-Zener tunneling or a renormalization of the electronic temperature \cite{CM:cdw,Han:cdw,Mazza:PRL,Matthies:PRB}. Another mechanism, experimentally confirmed in several cases \cite{VO2:PRB,Stefanovich_2000,VO2:heat}, is Joule heating to temperatures above the transition temperature.

However, the behavior of the current-driven transition observed in Ca$_2$RuO$_4$ appears to be inconsistent with these expectations \cite{Maeno:mainp,Fuku:CRO,Cao2:CRO,Dia:CRO,Xray:CRO,Cao:CRO,Friedt:CROPT,Pavarini:CRO,Cao2:CRO}. In this material the insulating phase is destroyed at threshold fields ($E_{\rm{th}}\sim40\um{V/cm}$) that are several orders of magnitude smaller than the fields required for a significant excitation of carriers, while it has been reported that the global temperature of the system remains below the equilibrium critical temperature \cite{Fuku:CRO}. Furthermore, a recent nano-imaging experiment \cite{Mengkun:CRO} observed coexistence of metallic and insulating phases when applying an increasing current to Ca$_2$RuO$_4$, and found that the metallic phase always nucleates out of the negative electrode, meaning that the phase switching depends on the direction of the current flow; another paper \cite{Cao2:CRO} reported a dependence of the hysteresis cycle on current direction. These findings indicate that Joule heating, which is proportional to the square of the current and hence does not change when the current polarity is reversed, is not the dominant effect.

In this work we reconsider heating effects in connection with metal-insulator transitions in correlated materials, with emphasis on a very simple point: in the presence of a current, a spatial variation in the Seebeck coefficient $S$ acts as a  heat source or a heat sink, depending on the direction of the current flow with respect to the Seebeck gradient. In the most common realization, a modulation in $S$ is produced by a temperature gradient ($\grad S\sim(dS/dT)\grad T$), while in a thermocouple the modulation occurs at a device boundary. Here we observe that in a system consisting of an inhomogeneous mixture of metal and insulating phases, similar effects may occur at interfaces between metallic and insulating phases, leading in appropriate cases to a marked dependence of the position of the phase boundaries on the direction of the current. We further show that because the difference in Seebeck coefficient between metal and insulating phases of correlated materials is typically large, the effect may be comparable to or larger than Joule heating. We  present a study of an idealized geometry that shows how these ideas may account for the essential features of the Ca$_2$RuO$_4$ data \cite{Mengkun:CRO}.

The rest of the paper is organized as follows. In Section \ref{sec:macroscopic} we use transport arguments to derive a heat balance equation that accounts for Joule heating, the Peltier (interface) effect, heat diffusion and heat dissipation into a reservoir. Section \ref{sec:Ca2RuO4} presents an analysis of  a specific geometry that is an idealized version of the Ca$_2$RuO$_4$ experiments. Section \ref{sec:Conclusions}  summarizes the results and outlines  directions for further research. Appendices provides a microscopic derivation of the heat balance based on the electronic quantum kinetic equation, and details of our estimations of experimental parameters.

\section{Heat balance equation}\label{sec:macroscopic}

We consider a system of electrons with electric charge $q=-e$, chemical potential $\mu$ and temperature $T$. We assume that a charge (electric) current density $\mb j_c$ exists in the system, along with an electric field $\mb E$. The relevant quantities are functions of the position $\mb x$, which we typically do not explicitly notate here. We write a steady state heat balance equation that relates $T$ to $\mb j_c$, beginning with consideration of the electronic energy density.

In the absence of magnetic fields, the total energy density $u_{tot}$ of the electronic system is the sum of the internal electron energy $u_{el}$ and of the electric field energy, and satisfies the equation of state $\mathrm{d}u_{tot}=\mathrm{d}u_{el}+\mb E\cdot\mathrm{d}\mb D/4\pi$, with $\mb D$ the displacement vector \cite{LL8}, implying that the continuity equation is
\begin{equation}\label{Ebal}
\partial_tu_{\rm{el}}+\frac{\mb E}{4\pi}\partial_t\mb D+\dive\mb j_e=-\dot Q_{\rm{d}},
\end{equation}
where $\mb j_e$ is the energy current and $\dot Q_{\rm{d}}$ describes the rate of energy dissipation into non-electronic degrees of freedom, such as the lattice modes in our case. The dissipation rate depends on the  temperature $T_l$ of the non-electronic degrees of freedom and the electronic temperature $T$: $\dot Q_{\rm{d}}=\dot Q_{\rm{d}}(T,T_l)$.

We use the fourth Maxwell equation to relate the displacement vector to the electrical current $\partial_t\mb D=-4\pi\mb j_c$, and introduce the electrochemical potential $\Phi$ via $-\grad\Phi\equiv\mb E-\grad\mu/q$. In a steady state $\partial_tu_{el}=0$ and $\dive\mb j_c=0$, we add and subtract $\dive(\mb j_c\mu/q)=\mb j_c\cdot\grad\mu/q$ in Eq. \eqref{Ebal} and introduce the heat current $\mb j_h=\mb j_e-\mu\mb j_c/q$ so that Eq. \eqref{Ebal} becomes
\begin{equation}\label{Hbal}
0=-\mb j_c\cdot\grad\Phi-\dive\mb j_h-\dot Q_{\rm d}.
\end{equation}
%
%
Equation \eqref{Hbal} represents the steady-state heat balance, relating the divergence of the heat current to Joule heating $-\mb j_c\cdot\grad\Phi$ and to heat dissipation into the reservoir.

We now use the linear theory of thermoelectric transport \cite{Kreuzer, Ashcroft} to write the heat and charge currents as functions of $-\nabla \Phi$ and of the electronic temperature gradient $\nabla T$. A standard form of this relation is
\begin{gather}\label{jcjs}
\begin{pmatrix}
\mb j_c \\
\mb j_h/T
\end{pmatrix}=\begin{pmatrix}
\sigma & \sigma S \\
\sigma S & \kappa_e/T+\sigma S^2
\end{pmatrix}\begin{pmatrix}
-\grad\Phi \\
-\grad T
\end{pmatrix},
\end{gather}
where $\sigma$ is the electric conductivity, $S$ the Seebeck coefficient (or thermopower) and $\kappa_e$ the electronic thermal conductivity. In Eq. \eqref{jcjs} the transport coefficients matrix is symmetric thanks to the Onsager relations. An equivalent form, based on charge and energy currents rather than charge and heat currents would involve the generalized forces $-\grad(\Phi/T)$ and $\grad(1/T)$ rather than $-\grad\Phi$ and $-\grad T$. The two formulations of course lead to equivalent results.

We now rearrange Eq. \eqref{jcjs} to obtain an expression for $\grad\Phi$ and $j_h$ in terms of $\mb j_c$ and $\grad T$:
\begin{equation}\label{Ejs}
\begin{pmatrix}
-\grad\Phi \\
\mb j_h
\end{pmatrix}=\begin{pmatrix}
\rho & -S \\
\Pi & \kappa_e
\end{pmatrix}\begin{pmatrix}
\mb j_c \\
-\grad T
\end{pmatrix},
\end{equation}
with $\Pi=TS$ the Peltier coefficient, $\rho=\sigma^{-1}$ the electrical resistivity and $\kappa_e$ is the thermal diffusion coefficient. Combining Eq. \eqref{Hbal} and \eqref{Ejs} and noting that the steady-state condition $\nabla\cdot\mb j_c=0$ implies $\dive(\Pi\mb j_c)=S\mb j_c\cdot\grad T+T\mb j_c\cdot\grad S$, yields
\begin{equation}\label{Heat}
0=-\dot Q_{\rm d}+\rho\mb j_c^2-T\mb j_c\cdot\grad S+\dive(\kappa_e\grad T).
\end{equation}
The first two terms in Eq. \eqref{Heat} are the dissipation into the reservoir and the Joule heating respectively, and the last term represents thermal diffusion. The remaining term, $T\mb j_c\cdot\grad S$ shows how spatial structure in the Seebeck coefficient in the presence of a current gives a thermal effect that may be either heating or cooling depending on the direction of current flow relative to the gradient of $S$. Heat is generated when the current flows from the phase with higher Seebeck coefficient to the phase with the lower one. In particular, a sharp interface separating two materials with  different thermoelectric coefficients is a localized heat source or sink.

Appendix ~\ref{Appendix:Derivation} gives a derivation of Eq. \eqref{Heat} from a microscopic approach, starting from the equations for the Keldysh Green functions and writing a kinetic equation for the electron distribution function.

Crucial to the solution of Eq. \eqref{Heat} is the dissipation of the generated heat into the thermal reservoir. In the situation of most interest here, the thermal reservoir is provided by the lattice degrees of freedom of the material. For simplicity, we assume that the heat transfer is proportional to an electron-lattice coupling $\gamma_{\rm{e-l}}$ (which for simplicity we take to be structureless) and to the difference in electron and lattice temperatures: $\dot Q_{\rm d}=\gamma_{\rm{e-l}}(T(\mb x)-T_l(\mb x))$. The heat balance equation for the lattice is then
\begin{equation}\label{HL1}
0=\gamma_{\rm{e-l}}(T-T_l)+\dive(\kappa_l\grad T_l),
\end{equation}
with $\kappa_l$ the lattice thermal conductivity. Note that there is neither Joule heating nor thermoelectric effects for the lattice. The heat that flows into the lattice will be dissipated into the environment, typically at the boundaries of the sample, leading to boundary conditions on $T_l$. A specific example will be discussed below.

Equation \eqref{HL1} implies that the length scale for variations in $T_l-T$ is $\sim \sqrt{\kappa_l/\gamma_{\rm{e-l}}}$. Typically this scale is $10-100\um{nm}$ (see Appendix \ref{Appendix:Ca2RuO4Parameters} for details) and is much shorter than the relevant geometrical length scales, so that to sufficient approximation $|T-T_l|\ll T_l\Rightarrow T_l\approx T$  and we can combine Eqs. \eqref{Heat} and \eqref{HL1} into an equation for just one temperature:
\begin{equation}\label{HT}
0=\rho\mb j_c^2-T\mb j_c\cdot\grad S+\dive(\kappa\grad T),
\end{equation}
with $\kappa\equiv\kappa_l+\kappa_e$ the total thermal conductivity, and with boundary conditions taken from those for Eq. \eqref{HL1}.

Equation \eqref{HT} determines the temperature, given the spatial dependence of $\rho$, $S$, $\kappa$ and the current. We suppose that the state of the system, and thus the values of the transport coefficients, only depends on the local temperature $T(\mb x)$. Given a certain dependence of the transport coefficients on $T(\mb x)$, then the continuity equation $\dive\mb j_c=0$, the third Maxwell equation $\rot\mb E=0=\rot(\rho\mb j_c)$ and Eq. \eqref{HT} completely determine the current and temperature. \footnote{Note that the boundary conditions on $\mb j_c$ and $\mb \grad T$ are that they vanish at the system surfaces, since no electrons can flow out of the system.}

\section{Results and Application to Ca$_2$RuO$_4$ }\label{sec:Ca2RuO4}

\subsection{Overview and parameters}

In this section we present analytical and numerical solutions of Eq. \eqref{HT}. When specific parameters are required we use values reasonable for Ca$_2$RuO$_4$.

Correlation-driven metal-insulator transitions are typically first order with narrow hysteresis regimes, and we assume that the transport coefficients take metallic or insulating values according to whether the local temperature is greater or less than the critical temperature $T_{\rm{MIT}}$. Thus the conductivity (resistivity) and thermopower take the values $\sigma_{M,I}$ ($\rho_{M,I}$) and $S_{M,I}$; we assume $\kappa_e\ll\kappa_l$ for simplicity so that $\kappa\approx\kappa_l$. Thus $\nabla S$ vanishes except at the metal-insulator phase boundaries, where it has delta-function singularities proportional to $\Delta S\equiv S_I-S_M$; this means that the Peltier heating is an interface effect while Joule heating is a bulk effect.

For  Ca$_2$RuO$_4$ at room temperature, the thermal conductivity is $\kappa\sim10^{-3}\um{W/cmK}$ (see Appendix \ref{Appendix:Ca2RuO4Parameters}), the insulating state resistivity is $\rho_I\sim10\um{\Omega cm}$ and the metal-insulator transition temperature is $T_{MIT}\approx 360K$; we define $\Delta T\equiv T_{\rm{MIT}}-T_0\sim60\um K$. The metal phase of Ca$_2$RuO$_4$ has a very low Seebeck coefficient while the insulating phase has a high and positive Seebeck coefficient \cite{Seeb:CRO}: $S_M\approx0$ and $S_I\approx400-1000\,\mu \rm{V/K}$; the sign of $S_I$ is consistent with the large particle-hole asymmetry found in dynamical mean field calculations \cite{QH:AJM,Ricco:CRO}. Thus in Ca$_2$RuO$_4$, $\grad S$ points from the metal to the insulating phase so the interface is heated when the current flows from the insulator to the metal, i.e. when the metal phase nucleates out of the negative electrode. This agrees with the experimental reports from Ref. \cite{Mengkun:CRO} regarding the dependence of the nucleation process on the direction of the electric current.

We study the geometry shown in Fig. \ref{IV}a, an idealized representation of a typical experimental geometry: a film of length $L$, width $W$ and thickness $h$ grown on a substrate, which we assume to be held at temperature $T_0$ and to act as a heat sink. We choose the $x$ axis to be along $L$, the $y$ axis to be along $W$ and the $z$ axis to be along $h$. Since for typical experiments $h\sim0.2\um{mm}\gg\sqrt{\kappa_l/\gamma_{\rm{e-l}}}$, we can use Eq. \eqref{HT} with boundary conditions $\partial_zT(z=0)=0$ and $T(z=h)=T_0$.

\subsection{Analytic Solutions}

Equation \eqref{HT} can be solved numerically, but analytical insight can be gained in particular limits. Although it is not directly relevant to most experimental situations, we assume $W\gg L\gg h\gg \sqrt{\frac{\kappa}{\gamma_{\rm{e-l}}}}$. We first consider a current $\mb j_c=j_0\hat x$ uniform in $x,y$, introduced at $x=0$ and removed at $x=L$. If the sample is entirely insulating the current does not depend $z$; the only source of heat is Joule heating and the temperature profile is given by (see Eqs. \eqref{TC1}-\eqref{TC3})
\begin{equation}\label{TIunif}
T(z)=T_0+\rho_Ij_0^2(h^2-z^2)/2\kappa.
\end{equation}

\begin{figure}[t]
\centering
\includegraphics[width=\columnwidth]{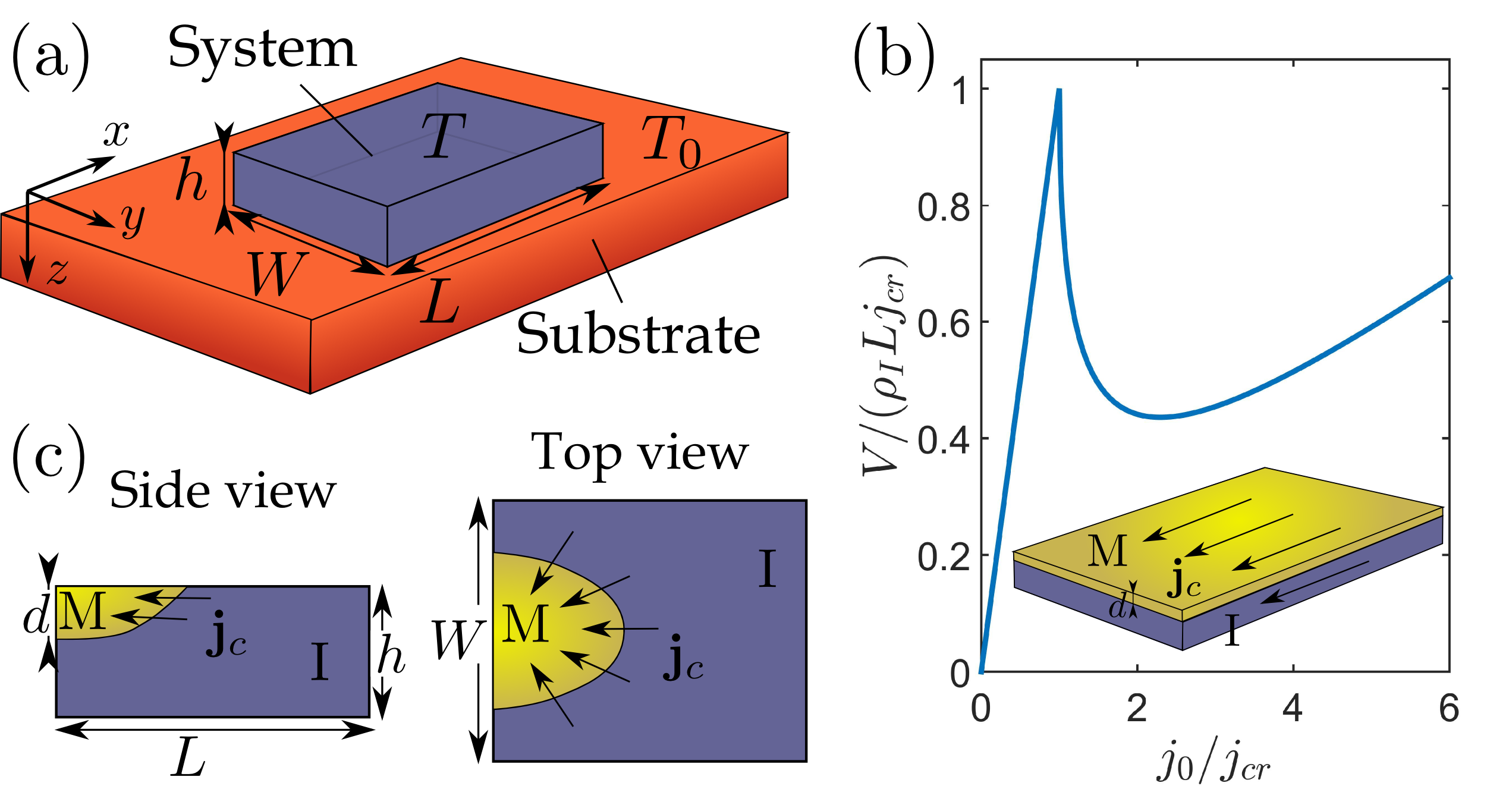}
\caption{\scriptsize{(a) Sketch of the geometry showing system (in blue) of length $L$, width $W$ and thickness $h$ and  substrate (red) which is held at temperature $T_0$. (b) Plot of the $I$-$V$ curve resulting from Eq. \eqref{d}; the cusp point is a consequence of the approximation used. The inset shows the geometry of the metallic and insulating phase. (c) Sketch of the side view and top view of the metallic and insulating phase in the situation considered for Eqs. \eqref{TPel} and \eqref{rc}.}}\label{IV}
\end{figure}

When $j_0$ reaches the critical current
\begin{equation}\label{jcrit}
j_{cr}\equiv\sqrt{2\Delta T\kappa/\rho_Ih^2}\sim5\um{A/cm^2},
\end{equation}
the temperature of the top surface reaches $T_{\rm{MIT}}$. For $j_0>j_{cr}$ a metallic region appears at the top surface. The estimate of $j_{cr}$ for Ca$_2$RuO$_4$ is in agreement with experimental reports \cite{Maeno:mainp,Fuku:CRO,Mengkun:CRO}. The appearance of a metallic phase near the top surface means that  the current becomes dependent on $z$. In the limit $h\ll L$, the region of $x$ in which $\mb j_c$ is not parallel to $\hat x$ is very small (of order $\sim h$), so that over most of the sample the current flows parallel to the interface and we can ignore any Peltier effect and focus on Joule heating only. In this geometry metal and insulator are essentially two resistors in parallel, so that the respective current densities $j_M$ and $j_I$ are related by $\rho_Ij_I=\rho_Mj_M$. Following Eqs. \eqref{jMjI}-\eqref{TdUnif} in Appendix \ref{Appendix:parallel current} we find the temperature at the interface $z=d$
\begin{equation}\label{Td}
T(d)=T_0+\Delta T\left(\frac{j_0}{j_{cr}}\right)^2\left[1-\frac{d^2}{(d+\frac{\rho_M}{\rho_I}(h-d))^2}\right].
\end{equation}
The condition $T(d)=T_{\rm{MIT}}$ determines the thickness of the metallic layer:
\begin{equation}\label{d}
d=h\frac{\rho_M/\rho_I\sqrt{1-j_{cr}^2/j_0^2}}{1-(1-\rho_M/\rho_I)\sqrt{1-j_{cr}^2/j_0^2}}.
\end{equation}
As expected $d(j_0=j_{cr})=0$, while $d$ approaches $h$ for $j_0\sim\sqrt{\rho_I/\rho_M}j_{cr}\gg j_{cr}$. From the value of $d$ we calculate the total current and find the V-I curve, shown in Fig. \ref{IV}b. The calculated curve exhibits the usual ``N'' shaped behavior expected for heating-driven insulator-metal transitions.

We now turn to an idealization of the situation studied in Ref. \cite{Mengkun:CRO}, in which a total current $I$ is injected through a point electrode at $x=L$, $y=W/2$, $z=0$ and removed at $x=0$, $y=W/2$, $z=0$. We focus on what happens near the $x=0$ electrode; in the limit $L\ll W$ and for points close to the electrode, the current decreases as $\mb j_c(r)\approx I/(\pi hr)\hat r$, with $r=\sqrt{x^2+(y-W/2)^2}$, with $I$ the total current. Crucially given the point like electrodes, all of the injected current must cross the insulator-metal phase boundary, and Peltier heating will play a role. The depth $d$ of the metallic region depends on $r$ and beyond the critical distance $r_c$ at which $d=0$ the sample is insulating at all $z$, as qualitatively sketched in Fig. \ref{IV}c.

We assume that over most of the relevant $r$ range in both metal and insulating regimes the current flow is perpendicular to $z$ and that the curvature in the $x-y$ plane of the interface may be neglected. Then the parallel resistor arguments imply that the total current in the metal (integrated along the $z$ direction) at distance $r$ is
\begin{equation}\label{IMt}
I_M(r)=\frac{d(r)}{d(r)+(h-d(r))\rho_M/\rho_I}\frac{I}{\pi r}.
\end{equation}
The current $I_M$ changes with $r$ thanks to the $r$ dependence of $d(r)$ so that some current must flow across the interface at $r$, giving rise to Peltier heating; this is a consequence of the point-like geometry of the electrodes which ensures an inhomogeneity in the $x-y$ plane.

We account for the Peltier contribution in the heat balance and require $T(d(r))=T_{\rm{MIT}}$, obtaining an equation for $\partial_rd$ which may be solved to find $d(r)$. If Joule heating is neglected we find (see Eqs. \eqref{IM}-\eqref{TdPH2} in Appendix \ref{Appendix:PeltierDetails})
\begin{equation}\label{TPel}
\Delta T=T_{\rm{MIT}}\Delta S\frac{I}{\pi r}\frac{\rho_M/\rho_I h^2}{\kappa(d+\rho_M/\rho_Ih)^2}\frac{\partial_rd}{\sqrt{1+(\partial_rd)^2}},
\end{equation}
where $\Delta S>0$. We observe that for $I<0$, i.e. current flowing into the metal, $\partial_rd$ is negative as it should be since $d(r=0)>0$ and $d(r_c>0)=0$; on the other hand, if $I>0$ then $\partial_rd\geq0$ and no stable interface can exist. In other words, for one direction of $I>0$ Peltier cooling contracts the interface to the very near vicinity of the electrode, while for the opposite direction it pushes the interface away from the electrode, consistently with observations in Ca$_2$RuO$_4$ that the metallic phase always nucleates out of the negative electrode, in such a way that the current flows from insulator into metal.

We solve Eq. \eqref{TPel} with $d(r=r_c)=0$ and integrating backwards to find $d(r)$. As explained in Appendix \ref{Appendix:PeltierDetails} (Eqs. \eqref{rd}-\eqref{rc2}), we can determine the critical radius by examining the solution for $r\rightarrow0$ under the assumption $d(r\rightarrow0)>\rho_M/\rho_Ih$ (corresponding to a larger current flow in the metal)
\begin{equation}\label{rc}
r_c\approx\sqrt{2hT_{\rm{MIT}}|I|\Delta S/\pi\kappa\Delta T}.
\end{equation}
Note that the metal phase disappears into the insulator at a nonzero angle, since $\partial_rd$ is nonzero for $r\rightarrow r_c$. This is in agreement with the wedge-shaped metallic phase considered in the elastic theory of stripe formation in Ref. \cite{Mengkun:CRO}. For $I\sim10\um{mA}$ and parameters compatible with Ca$_2$RuO$_4$ we find $r_c\sim0.2\um{mm}$ which is a sizable fraction of the length $L\sim1\um{mm}$; we thus show that even in a very simple limit our model reproduces the qualitative experimental features reported in Ref. \cite{Mengkun:CRO}.
\begin{figure*}[t]
\centering
\includegraphics[width=\textwidth]{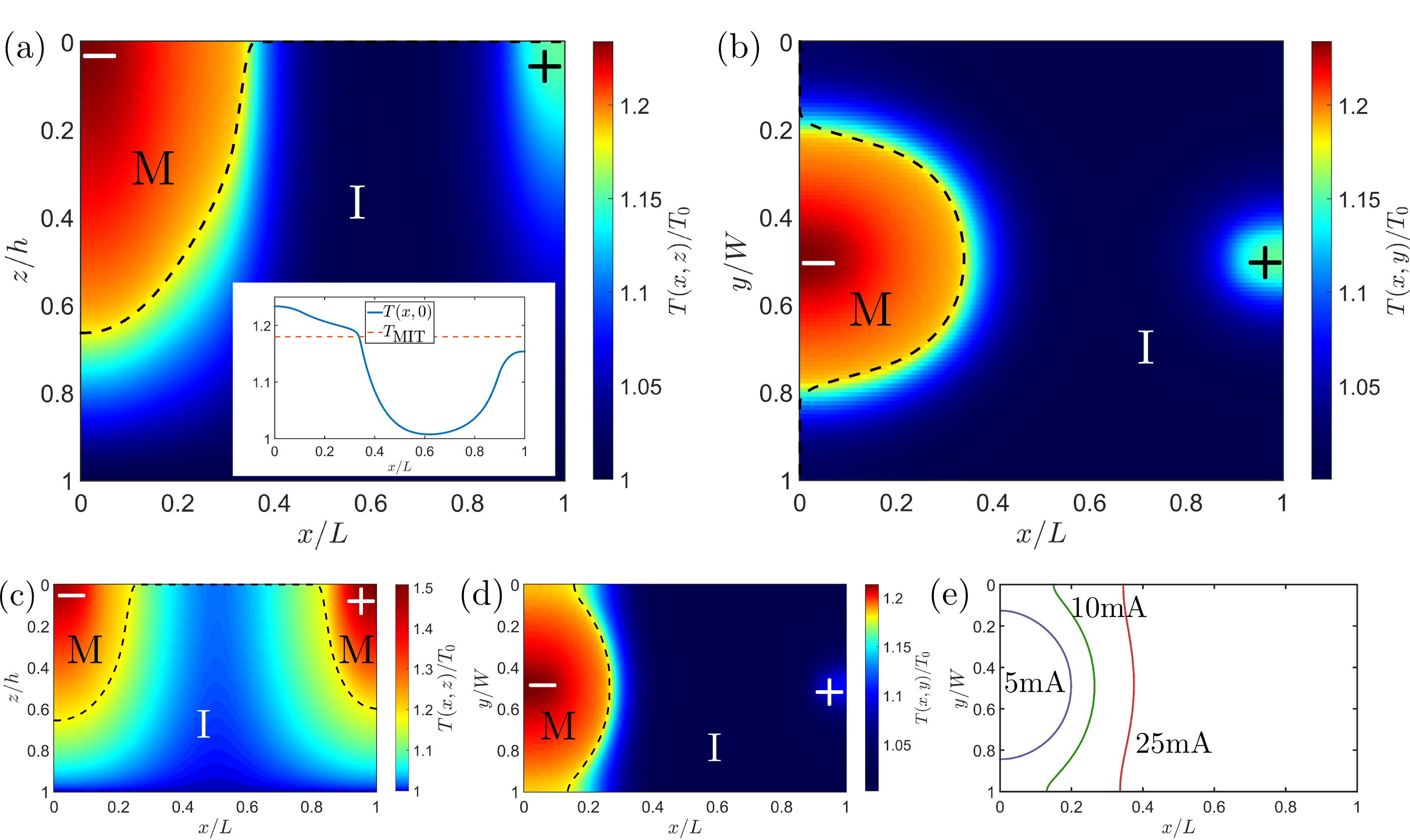}
\caption{\scriptsize{(a) Color map of the temperature $T(x,z)$ evaluated at $y=W/2$ and normalized to the external temperature $T_0$ for $h/L=0.2$, $W=L$, $\rho_M/\rho_I=0.1$, $\rho_II^2/(\kappa L^2T_0)=0.2$, $\rho_M/\rho_I=0.1$, $I\Delta S/(\kappa L)=-0.2$, with $I\sim10\um{mA}$ the total current; these parameters correspond to a realistic simulation of Ca$_2$RuO$_4$ experiments. The negative/positive electrode is located at $(0<x<0.1,y=W/2,z=0)$/$(0.9<x<1,y=W/2,z=0)$; the dashed black line shows the trial phase boundary used, the $M/I$ labels indicate the metal/insulator phase. The inset shows the plot of $T(x,y=W/2,z=0)/T_0$ as function of $x$ compared to the transition temperature $T_{\rm{MIT}}$. (b) Color map of $T(x,y,z=0)/T_0$ for the parameters of (a). (c) Color map of $T(x,y=W/2,z)/T_0$ for the same parameters as (a), except for a small Seebeck coefficient $I\Delta S/(\kappa L)=-0.02$ and a larger current $\rho_II^2/(\kappa L^2T_0)=1.2$; from the approximate symmetry of the metal phase we can see that Joule heating is the dominant effect. (d) Color map of $T(x,y,z=0)/T_0$ for the same parameters of (a) and (b), but $W=L/2$; in a rectangular geometry the metal phase can extend from $y=0$ to $y=W/2$. (e) Shape of the phase boundary for the parameters of (d) and different values of the current $I=5,10,25\um{mA}$.}}\label{T2D}
\end{figure*}

\subsection{Numerical Solution}

We now consider both Joule heating and Peltier effect, and iteratively solved the coupled system given by $\dive\mb{j}_c=0$, $\rot(\rho\mb{j}_c)=0$ and Eq. \eqref{HT}, using the parameters listed above which are appropriate to Ca$_2$RuO$_4$. We assume that the transport coefficients are determined by the local temperature $T(\mb x)$, i.e. $\sigma(T(\mb x)>T_{\rm{MIT}})=\sigma_M$ and $\sigma(T(\mb x)<T_{\rm{MIT}})=\sigma_I$. The system has dimensions $L=1\um{mm}$, $W\sim L$ and $h/L=0.2$ and the negative (positive) electrode is located at $0<x/L<0.1$ ($0.9<x/L<1$), $y=W/2$, $z=0$.

We present in Fig. \ref{T2D} two different cases, for the point electrodes geometry. Panels (a) and (b) shows computations for $S_I\sim1000\um{\mu V/K}$,  which is a value representative of Ca$_2$RuO$_4$, and a square geometry $W=L$; the other parameters are given in the figure caption. The simulation clearly shows that a current $I\sim10\um{mA}$ can induce a metal phase extending over part of the sample (up to $x/L\sim0.3$; see inset in Fig. 2a). Note also that the metal phase exists only on one side of the sample, showing that the interface Peltier effect leads to a pronounced spatial asymmetry of the metallic domains, consistent with observations of Ref. \cite{Mengkun:CRO}. Panel (b) shows the top view of the surface temperature $T(x,y)$; we see that for distances from the electrode less than $\sim 0.3L$ the temperature is above the metal-insulator transition temperature, while for the rest of the sample the temperature is below. Panel (c) shows results of analogous computations, in which the Seebeck coefficient is reduced to a much smaller value, showing that in this case two metallic phases nucleate around both electrodes in a nearly symmetric way; this occurs when the Joule heating plays is more relevant and the effect is thus almost symmetric in the direction of the current. These results show that Peltier heating at interfaces can explain the appearance of a stable metal phase in typical Ca$_2$RuO$_4$ experiments even for modest values of the applied current.

We notice that the thickness $d$ of the metallic layer is a sizeable fraction of $h$, and depends on the value of $S$ and on of the resistivity ratio $\rho_I/\rho_M$; in particular $d$ decreases for larger $\rho_I/\rho_M$. Notice that since $h$ is not much smaller than $L$ as assumed in the derivation of Eq. \eqref{TPel}, the boundary geometry is only qualitatively a wedge.

In Fig. 2d and 2e we consider a rectangular geometry $W=L/2$, observing that the maximum temperature does not change appreciably, while the interface geometry is modified; in Fig. 2e we study the evolution of the phase boundary for increasing currents and find that it extends to larger values of $x/L$, in agreement with the results of Ref. \cite{Mengkun:CRO}, and can stretch along the entire width of the sample. We also considered the case of a wider geometry $W=2L$ and found that the phase boundary has a geometry similar to that of the square $W=L$ case and is more stretched along the $x$ axis rather than along the $y$ axis.

We can also calculate the average temperature of the system; for the parameters of Fig. 2a we find an average temperature $\bar T=1.05T_0$ and an average top surface temperature $\bar T(z=0)=1.07T_0$. This shows that even for modest increases of the global temperature of the system, $T$ can be locally larger than the critical temperature.


\section{Conclusions}\label{sec:Conclusions}

We considered a correlated system with an electric current flowing through an  interface between its metallic and insulating phases.  Using macroscopic arguments based on entropy production and linear transport theory, we wrote a heat balance equation that can be solved for the temperature given the current. The equation accounts for Joule heating, heat diffusion, heat dissipation, and crucially, includes the Peltier effect, which linearly couples the current to the discontinuity $\Delta S$ in thermopower across an interface between a metal and an insulator. This term leads to heating or cooling at the interface depending on the direction of the current with respect to the change in thermopower. The magnitude of the interface Peltier effect depends on the current, the thermopower change across the interface, and the efficiency of heat dissipation (determined here by the thermal conductivity and the distance to the nearest heat sink). Because the Peltier heating is linear in the current while Joule heating is quadratic, the Peltier heating dominates at small currents, and leads to interesting physics if the Peltier effect remains dominant at currents large enough to heat the system above its insulator-metal transition temperature. This condition is equivalent to a thermoelectric-like figure of merit $ZT\equiv T\Delta S^2/\rho\kappa$ being larger than one. Our analysis showed that for Ca$_2$RuO$_4$ in geometries similar to those considered in recent experiments \cite{Mengkun:CRO}, heating effects can stabilize a non-equilibrium metal phase in a region of the sample that depends on the direction of the current. Other materials that exhibit a large discontinuity in thermopower at the metal insulator transition and may also exhibit a similar polarity dependence include VO$_2$ microbeams \cite{Seeb:VO2} and Cu$_2$Se \cite{Seeb:Cu2Se}.

Interesting directions for future theoretical research include extension of our analysis to other experimental geometries, in particular to the very thin sample regime and to the filamentary conduction pattern observed in other correlated electron systems.  Experimental observations of  our predicted distribution of the local temperature of the sample surface would be of great interest. A very recent work \cite{Mattoni:CRO} by Mattoni et al. studied the local temperature in Ca$_2$RuO$_4$ under current and reported data compatible with our predictions. Equally of great interest would be the investigation of different electrode geometries (e.g. introducing the current uniformly across a sample face rather than in a small region) and sample thicknesses. Finally, our analysis is primarily based on a macroscopic theory, and a more detailed microscopic analysis of the physics of the metal-insulator interface, in the presence of current flow, would be of considerable interest.

\textit{Acknowledgements} We thank A. Georges, A. McLeod, and, particularly, Mengkun Liu for very helpful discussions. This work was supported by the US Department of Energy under grant DE-SC0012375.

\bibliographystyle{apsrev4-1}
\bibliography{Polarity-bib}

\onecolumngrid

\pagebreak[4]
\appendix

\section{Derivation from microscopics}\label{Appendix:Derivation}

In this Appendix we derive the kinetic equations from  a theory of electrons subject to various forms of scattering. We consider the electrons in a quasi-particle picture, so they are characterized by their charge $q=-e$, the momentum $\vk$ and a generally position dependent energy $\epsilon_{\vk}(\mb x)$.

From the equations of motion for the Keldysh Green functions, we write down a kinetic equation for the lesser component $G^<_{\vk}(\varepsilon,\mb x,t)$ \cite{Keld,JH:book,Kam:book}:
\begin{equation}\label{Gl1}
i(\partial_t+\mb v_{\vk}\cdot\grad-\grad\epsilon_{\vk}\cdot\grad_{\vk}+q\mb E\cdot(\grad_{\vk}+\mb v_{\vk}\partial_{\varepsilon}))G^<_{\vk}(\varepsilon,\mb x,t)=\St{tr}\{G^<_{\vk}\}+\St{in}\{G^<_{\vk}\}.
\end{equation}
Here $\grad$ is the gradient in real space, $\grad_{\vk}$ is the gradient in momentum space, $\mb v_{\vk}\equiv\grad_{\vk}\epsilon_{\vk}$ is the electron velocity, $\mb E$ is the electric field, $\St{tr}$ is the transport collision integral and $\St{in}$ is the energy relaxation collision integral. The second and third driving term in the LHS of Eq. \eqref{Gl1} come from the space inhomogeneity, while the last two terms describe the effect of the electric field. From symmetry consideration, we observe that the transport scattering relaxes momentum, so it must be odd in $\vk$, i.e. $\sum_{\vk}\St{tr}\{G^<\}=0$, while we assume that $\St{in}$ does not depend on momentum but only on $\varepsilon$.

The electric current $\mb j_c$ and the energy current $\mb j_e$ can be written as
\begin{equation}\label{jcjeK}
\mb j_c(\mb x,t)=q\int\frac{d\varepsilon}{2\pi i}\sum_{\vk}\mb v_{\vk}G^<_{\vk}(\varepsilon,\mb x,t);\qquad \mb j_e(\mb x,t)=\int\frac{d\varepsilon}{2\pi i}\sum_{\vk}\epsilon_{\vk}\mb v_{\vk}G^<_{\vk}(\varepsilon,\mb x,t).
\end{equation}

We work in the small currents limit, so that away from the interface the electric field is small enough to not change in any significant way the retarded part of the Green function, which is then completely determined by the equilibrium band structure. However, at the interface huge fields that break this assumption may be generated by the change in the electronic structure; for the sake of simplicity, we consider a very abrupt phase interface, with thickness going to zero, so that we can operate under the small fields assumption at every $\mb x$. This assumption is reasonable, since the typical thickness of domain walls is usually a few lattice constants, while the scales associated to transport are bigger.

The imaginary part of the retarded Green function is then proportional to the spectral weight $A_{\vk}(\varepsilon,\mb x,t)$. We do not need to specify the exact form of $A_{\vk}$, but simply to assume that it is a peaked function of $\varepsilon-\epsilon_{\vk}(\mb x)$; the spectral weight is related to the density of states by $2\pi iD(\varepsilon,\mb x)\equiv\sum_{\vk}A_{\vk}(\varepsilon, \mb x)$.

We now use the generalized Kadanoff-Baym ansatz and write $G^<_{\vk}$ as a momentum symmetric part plus a momentum anisotropic term $\delta G_{\vk}$ arising from the presence of the electric field:
\begin{equation}\label{GKBA}
G_{\vk}^<(\varepsilon,\mb x)\approx A_{\vk}(\varepsilon,\mb x)f(\varepsilon,\mb x)+\delta G_{\vk};\quad \sum_{\vk}\delta G_{\vk}=0;\quad
\sum_{\vk}G_{\vk}^<\equiv\bar G^<(\varepsilon,\mb x);\quad \sum_{\vk}\mb v_{\vk}G_{\vk}^<=\sum_{\vk}\mb v_{\vk}\delta G_{\vk}\equiv\mb F,
\end{equation}
where $f(\varepsilon,\mb x)$ is the electron distribution. In other words, $\bar G^<$ represents the part of $G_{\vk}^<$ that is even in momentum, while $\mb F$ represents the $\vk$-odd part of $G_{\vk}^<$.

We can now apply $\sum_{\vk}$ and $\sum_{\vk}\mb v_{\vk}$ to Eq. \eqref{Gl1} and use Eq. \eqref{GKBA} to get two equations for $\bar G^<$ and $\mb F$ from which we can obtain a kinetic equation for $f$ and an expression for the currents in terms of $f$.

We start by applying $\sum_{\vk}$ to Eq. \eqref{Gl1}. We use the identity $\sum_{\vk}(\mb v_{\vk}\cdot\grad G_{\vk}^<-\grad\epsilon_{\mb k}\cdot\grad_{\mb k}G_{\mb k}^<)=\sum_{\vk}[\grad\cdot(\mb v_{\mb k}G_{\mb k}^<)-\grad_{\mb k}\cdot(\grad\epsilon_{\mb k}G_{\mb k}^<)-(\grad\cdot\mb v_{\mb k}-\grad_{\mb k}\cdot\grad\epsilon_{\mb k})G_{\mb k}^<]=\grad\cdot\sum_{\vk}\mb v_{\mb k}G_{\mb k}^<-\sum_{\vk}\grad_{\vk}(\grad\epsilon_{\vk}G^<_{\vk})=\dive\mb F$, where sums over $\vk$ of total derivatives with respect to $\vk$ vanish, and obtain
\begin{equation}\label{dtG}
i\partial_t\bar G^<+i(q\mb E\partial_{\varepsilon}+\grad)\cdot\mb F=\sum_{\vk}\St{in}\{G^<_{\vk}\}
\end{equation}

We now apply $\sum_{\vk}\mb v_{\vk}$ to Eq. \eqref{Gl1} and write the transport scattering in a time relaxation approximation $\sum_{\vk}\mb v_{\vk}\St{tr}\{G_{\vk}^<\}\approx-i\sum_{\vk}\mb v_{\vk}G_{\vk}^</\tau_{\rm{tr}}$, so that
\begin{equation}\label{dtF}
i\partial_t\mb F+i\sum_{\mb k}\mb v_{\mb k}[\mb v_{\mb k}\cdot\nabla -\nabla\epsilon_{\mb k}\cdot\nabla_{\mb k}+q\mb E\cdot(\nabla_{\mb k}+\mb v_{\mb k}\partial_{\varepsilon})]G_{\mb k}^<=-i\frac{\mb F}{\tau_{\rm{tr}}}
\end{equation}

We now assume that the transport scattering is much stronger than the effect of the deviations from equilibrium, i.e. $qE\tau_{\rm{tr}}\ll\hbar/a$ (with $a$ the typical lattice cell size). In this limit, the effect of the electric field and of the space inhomogeneity on the Green functions is very small and the $\vk$-anisotropy satisfies $\delta G_{\vk}^<\ll A_{\vk}f$. We therefore replace $G_{\vk}^<$ with $A_{\vk}f$ in Eq. \eqref{dtF} and neglect $\partial_t\mb F\ll\mb F/\tau_{\rm{tr}}$, writing
\begin{equation}\label{F}
\mb F\approx-\sum_{\vk}\tau_{\rm{tr}}A_{\vk}\mb v_{\vk}\mb v_{\vk}\cdot(\grad+q\mb E\partial_{\varepsilon})f=-2\pi i\tau_{\rm{tr}}D(\varepsilon)\mb{v^2}(\varepsilon)(\grad+q\mb E\partial_{\varepsilon})f(\varepsilon).
\end{equation}
We have used that $A_{\vk}(\varepsilon,\mb x)=A(\varepsilon-\epsilon_{\vk}(\mb x))$ and thus $(\mb v_{\vk}\partial_{\varepsilon}+\grad_{\mb k})A_{\mb k}(\varepsilon,\mb x)=(\mb v_{\vk}\partial_{\varepsilon}-\grad_{\mb k}\epsilon_{\mb k}\partial_{\varepsilon})A_{\mb k}(\varepsilon,\mb x)=0$ and $(\mb v_{\mb k}\cdot\grad -\grad\epsilon_{\mb k}\cdot\grad_{\mb k})A_{\mb k}(\varepsilon,\mb x)=(-\mb v_{\mb k}\cdot\grad\mb\epsilon_{\mb k}\partial_{\varepsilon}+\grad\epsilon_{\mb k}\cdot\grad_{\mb k}\epsilon_{\mb k}\partial_{\varepsilon})A_{\mb k}(\varepsilon,\mb x)=0$. The tensor $\mb{v^2}$ is defined by $2\pi iD(\varepsilon,\mb x)(\mb{v^2})^{ab}(\varepsilon,\mb x)\equiv\sum_{\vk}A_{\vk}v^a_{\vk}v^b_{\vk}$. We can now substitute Eq. \eqref{F} into Eq. \eqref{jcjeK} to obtain an expression for the currents and into Eq. \eqref{dtG} to get a kinetic equation for $f$:
\begin{gather}\label{jcje}
\mb j_c(\mb x)=-q\int d\varepsilon\tau_{\rm{tr}}D(\varepsilon)\mb{v^2}(\varepsilon)(\grad+q\mb E\partial_{\varepsilon})f(\varepsilon,\mb x); \qquad
\mb j_e(\mb x)=-\int d\varepsilon\varepsilon\tau_{\rm{tr}}D(\varepsilon)\mb{v^2}(\varepsilon)(\grad+q\mb E\partial_{\varepsilon})f(\varepsilon,\mb x);\\
\label{Kef}\partial_t(D(\varepsilon,\mb x,t)f(\varepsilon,\mb x,t))-(\grad+q\mb E\partial_{\varepsilon})\tau_{\rm{tr}}D(\varepsilon,\mb x,t)\mb{v^2}(\varepsilon,\mb x,t)(\grad+q\mb E\partial_{\varepsilon})f(\varepsilon,\mb x,t)=\St{f}\{f\},
\end{gather}
where we have used $i\partial_t\bar G^<=-2\pi\partial_t(Df)$ and defined $\St{f}\equiv-\frac1{2\pi}\sum_{\vk}\St{in}\{G_{\vk}^<\}$. Since we consider a steady state situation we will set $\partial_t(Df)=0$ from now on.

Equations \eqref{jcje}-\eqref{Kef} completely determine the transport properties of the system. In order to make a connection to Eq. \eqref{jcjs}, we need to define a chemical potential $\mu(\mb x)$ and a temperature $T(\mb x)$; we observe that we can always write the $\mb x$ dependence of the distribution as $f=f(\varepsilon-\mu(\mb x), T(\mb x))$. In fact, the chemical potential is determined by the Poisson equation, since it expresses the electron density balance, and the temperature is given by the solution of the kinetic equation \eqref{Kef}. In this limit we can write $\grad f=-\grad\mu\partial_{\varepsilon}f+\grad T\partial_Tf$ and thus express the currents $\mb j_c$ and $\mb j_s=(\mb j_e-\mu\mb j_c)/T$ in Eq.\eqref{jcjs} as linear functions of $\mb E-\grad\mu/q$ and $\grad T$. In particular the transport coefficients are
\begin{equation}\label{SP}
\sigma S=e\int d\varepsilon\tau_{\rm{tr}}D\mb{v^2}(-\partial_Tf);\quad
\sigma \Pi=e\int d\varepsilon\tau_{\rm{tr}}D\mb{v^2}(\varepsilon-\mu)\partial_{\varepsilon}f;\quad
\sigma=e^2\int d\varepsilon\tau_{\rm{tr}}D\mb{v^2}(-\partial_{\varepsilon}f),
\end{equation}
where we used $q=-e$.

We observe that the Onsager relation $\Pi(\mb x)=T(\mb x)S(\mb x)$ is not immediate from Eq. \eqref{SP}, since it is not guaranteed that $-\partial_Tf=\frac{\varepsilon-\mu}T\partial_{\varepsilon}f$. However, the Onsager relations hold in the linear response regime. For coefficients calculated in the absence of external fields; it is possible to show that the Onsager relation is indeed satisfied by Eq. \eqref{SP} in such regime. The derivation requires some general properties of the energy relaxation collision integral $\St{f}$. Such quantity does not depends on $\mb x$, because the scattering processes occur on a local scale, and vanishes when $f$ is a Fermi-Dirac distribution, i.e. when local thermal equilibrium has been attained between the electrons or with a reservoir, depending on the scattering mechanism. Since the LHS of Eq. \eqref{Kef} is at least first order in the driving fields, the distribution at the zeroth order must be a thermal distribution $f=[e^{\frac{\varepsilon-\mu(\mb x)}{T(\mb x)}}+1]^{-1}+\mathcal O(\mb E-\grad\mu/q,\grad T)$. Such distribution satisfies $-\partial_Tf=\frac{\varepsilon-\mu}T\partial_{\varepsilon}f$ and thus $\Pi(\mb x)=T(\mb x)S(\mb x)$.

From Eqs. \eqref{jcje}-\eqref{Kef} we can finally derive a heat balance equation by applying $\int d\varepsilon\varepsilon$ to Eq. \eqref{Kef} and finding
\begin{equation}\label{HK}
-\grad\int d\varepsilon\varepsilon\tau_{\rm{tr}}\mb{v^2}(\grad+q\mb E\partial_{\varepsilon})f+q\mb E\int d\varepsilon\tau_{\rm{tr}}\mb{v^2}(\grad+q\mb E\partial_{\varepsilon})f=\int d\varepsilon\varepsilon\St{f}\{f\};\quad\Rightarrow\dive\mb j_e-\mb E\cdot j_c=-\dot Q_{\rm d},
\end{equation}
where we have defined $-\dot Q_{\rm d}\equiv\int d\varepsilon\varepsilon\St{f}\{f\}$. Equation \eqref{HK} is exactly Eq. \eqref{Ebal}, showing the macroscopic approach is fully consistent with a microscopic treatment of the problem.

To conclude we show that the dissipated heat is approximately linear in the temperature difference when the inelastic scattering is given by energy diffusion into a thermal bath at temperature $T_l$. In this case, the collision integral is written as \cite{CM:cdw,CMA:neg} $\St{f}\sim\partial_{\varepsilon}[D^2(\varepsilon)(T_l\partial_{\varepsilon}f+f(1-f))]$ and thus
\begin{equation}\label{Qd}
-\dot Q_{\rm d}=\int d\varepsilon\varepsilon\St{f}=-\Gamma\int d\varepsilon D^2(-T_l/T+1)f(1-f)\approx-\gamma_{\rm{e-l}}(T-T_l),
\end{equation}
where we used that for a thermal distribution $\partial_{\varepsilon}f=-\frac1Tf(1-f)$ and $\int f(1-f)=T$ and defined $\gamma_{\rm{e-l}}\equiv\Gamma D(\mu)^2$. Equation \eqref{Qd} shows that the assumption made in the main text of a linear dependence on $T$ of the dissipated heat is reasonable; furthermore, the dependence of $\gamma_{\rm{e-l}}$ on $T$ is small (it is of order $T^2$), while the dependence on $\mu$ arises from $D(\mu)$ and can usually be neglected compared to the linear dependence on $T$ of the $T-T_l$ term.

\section{Estimation of parameters}\label{Appendix:Ca2RuO4Parameters}

\subsection{Seebeck coefficient}

Using Eq. \eqref{SP}, we write the Seebeck coefficient as
\begin{equation}\label{Sins}
S=\frac1e\frac1T\left(\int d\varepsilon \tau_{\rm{tr}}D\mb{v^2}(\varepsilon-\mu)\partial_{\varepsilon}f\right)\left(\int d\varepsilon \tau_{\rm{tr}}D\mb{v^2}(-\partial_{\varepsilon}f)\right)^{-1}.
\end{equation}

We can calculate this for an insulating phase with a gap $\Delta$, so that the valence band extends up to $\varepsilon=0$ and the conduction band extends from $\varepsilon=\Delta$ up. The chemical potential is $\mu=\frac{\Delta}2+\frac T2\ln\left(\frac{\bar D_V}{\bar D_C}\right)$, where the effective densities of states are given by $\bar D_C\equiv\int_0^{\infty}D(\Delta+xT)e^{-x}dx$ and $\bar D_V\equiv\int_0^{\infty}D(-xT)e^{-x}dx$. In this approximation, and assuming the the energy dependence of $\tau_{\rm{tr}}\mb{v^2}$ is small, we find that the Seebeck coefficient is approximated by
\begin{equation}\label{Sfin}
S=\frac{k_B}{2e}\ln\left(\frac{\bar D_V}{\bar D_C}\right),
\end{equation}
where we have restored the Boltzmann constant $k_B$. A bigger spectral weight of holes compared to electrons leads to $\bar D_V>\bar D_C$ and thus to a positive Seebeck coefficient. This is indeed the case for Ca$_2$RuO$_4$, as confirmed by direct experimental measurements of $S$ \cite{Seeb:CRO} and by calculation and measurements of the density of states \cite{QH:AJM,Ricco:CRO}.

\subsection{Estimate of $\kappa$}

In this subsection we analyze Eqs. \eqref{Heat} and \eqref{HL1} and determine in which limit Eq. \eqref{HT} is applicable. We also estimate the total thermal conductivity $\kappa=\kappa_e+\kappa_l$ for Ca$_2$RuO$_4$.

We assume constant thermal conductivities and no dependence on $x$ and $y$ coordinates, but only on $z$; we also assume constant current $|\mb j_c|=j$ and no phase interface for simplicity. We find the electron temperature from Eq. \eqref{HL1} and substitute into Eq. \eqref{Heat}:
\begin{equation}\label{TC1}
T(z)=T_l(z)-\frac{\kappa_l}{\gamma_{\rm{e-l}}}\partial_z^2T_l;\qquad
-\frac{\kappa_e\kappa_l}{\gamma_{\rm{e-l}}}\partial_z^4T_l+\kappa\partial_z^2T_l+\rho j^2=0.
\end{equation}

The fourth order differential equation for $T_l$ has a solution of the form
\begin{equation}\label{TC2}
T_l(z)=A+Bz-\frac{\rho j^2}{2\kappa}z^2+Ce^{-\alpha z}+De^{\alpha z};\qquad\alpha^2=\gamma_{\rm{e-l}}\frac{\kappa}{\kappa_e\kappa_l},
\end{equation}
where the constants are determined by the boundary conditions $T_l(h)=T_0$, $\partial_zT_l(0)=0$, $\partial_zT(0)=0$ and $\partial_zT(h)=0$. We obtain
\begin{equation}\label{TLsol}
T_l(z)=T_0+\frac{\rho j^2}{2\kappa}(h^2-z^2)+\frac{\rho j^2}{\kappa\alpha h \sinh(\alpha h)}[\cosh(\alpha h)-\cosh(\alpha z)]
\end{equation}

Typical values of the electronic dissipation rate are $\gamma_{\rm{e-l}}\sim10^7\um{W/cm^3K}$, while the conductivities are typically $\kappa_l\sim10^{-3}\um{W/cmK}$ and $\kappa_e\lesssim\kappa_l/10$, so that $\alpha\sim10^5\um{cm^{-1}}$ and for single crystal systems $\alpha h\gg1$. The last term in Eq. \eqref{TLsol} is thus negligible and from Eq. \eqref{TC1} we find $T-T_l=\kappa_l/\gamma_{\rm{e-l}}\partial_z^2T_l\sim(T_l-T_0)\kappa_l/(\gamma_{\rm{e-l}}h^2)(T_l-T_0)$; again $\kappa_l/(\gamma_{\rm{e-l}}h^2)\ll1$ for $h\gg100\um{nm}$, so that $|T-T_l|\ll|T_l-T_0|\lesssim T_l$ for most single crystal systems (thin films may be thinner and break this regime). Within this approximation we combine Eq. \eqref{Heat} and \eqref{HL1} into Eq. \eqref{HT} as shown in the main text and write the solution for the electronic temperature in the case of constant current and resistivity as
\begin{equation}\label{TC3}
T(z)\approx T_l(z)=T_0+\frac{\rho j^2}{2\kappa}(h^2-z^2).
\end{equation}

In particular, the lattice temperature measured on the top surface of the system is $T_l(0)=T_0+\frac{\rho j^2}{2\kappa}h^2$. From this we estimate $\kappa$ for Ca$_2$RuO$_4$ using the data available in Ref. \cite{Fuku:CRO}: $h=0.2\um{mm}$, $T_0=273\um K$, $T_l(h,j=20\um{A/cm^2})\approx303\um K$, $\rho(j=20\um{A/cm^2})=0.5\um{\Omega cm}$, so that $\kappa\approx1.3\cdot10^{-3}\um{W/cmK}$. It is also possible to estimate the electronic thermal conductivity and find that it is much smaller than $\kappa_l$, so that $\kappa\approx\kappa_l$.

If the system is completely insulating, then $\rho=\rho_I$ and Eq. \eqref{TC3} gives Eq. \eqref{TIunif}.

\section{Solution of heat balance equation in parallel current geometry \label{Appendix:parallel current}}

We now consider the situation in the main text with a metal extending in $0<z<d$ and the insulator in $d<z<h$ and with total current density $j_0$. The current densities in the metal $j_M$ and in the insulator are given by
\begin{equation}\label{jMjI}
jMd+j_I(h-d)=j_0h;\qquad \rho_Mj_M=\rho_Ij_I\,\,\Rightarrow\,\,j_M=\frac{h}{d+(h-d)\rho_M/\rho_I}j_0,
\end{equation}
so that Eq. \eqref{HT} becomes
\begin{equation}\label{HTjMjI}
\partial_z^2T=-\frac{\rho_M j_0^2}{2\kappa}\left(\frac{h}{d+(h-d)\rho_M/\rho_I}\right)^2\,\,
\begin{cases}
      1 & 0<z<d\\
      \rho_M/\rho_I & d<z<h
    \end{cases},
\end{equation}
with boundary conditions $\partial_zT(0)=0$ and $T(h)=0$. Integrating Eq. \eqref{HTjMjI} from the bottom, we find $T(z)$ for $d<z<h$
\begin{equation}\label{TsolUnif}
T(z)=T_0+\frac{\rho_M j_0^2}{2\kappa}\left(\frac{h}{d+(h-d)\rho_M/\rho_I}\right)^2\frac{\rho_M}{\rho_I}(h^2-z^2)+\frac{\rho_M j_0^2}{\kappa}\left(\frac{h}{d+(h-d)\rho_M/\rho_I}\right)^2d(h-z)(1-\rho_M/\rho_I),
\end{equation}
and evaluating it at $z=d$ we obtain Eq. \eqref{Td}
\begin{gather}
\notag T(d)=T_0+\frac{\rho_I j_0^2}{2\kappa}\left(\frac{h}{d+(h-d)\rho_M/\rho_I}\right)^2\left[(h^2-d^2)(\rho_M/\rho_I)^2+
2d(h-d)\rho_M/\rho_I(1-\rho_M/\rho_I)\right];\\
\label{TdUnif}
T(d)=T_0+\frac{\rho_I j_0^2h^2}{2\kappa}\frac{(d+(h-d)\rho_M/\rho_I)^2-d^2}{(d+(h-d)\rho_M/\rho_I)^2}=T_0+\Delta T\left(\frac{j_0}{j_{cr}}\right)^2\left[1-\frac{d^2}{(d+(h-d)\rho_M/\rho_I)^2}\right].
\end{gather}

\section{Heat balance equation in the absence of Joule heating}\label{Appendix:PeltierDetails}

%
We now consider the situation of a current $I$ being injected at $x=0,z=0$ and a metal phase extending in $0<z<d(r)$, as explained in the main text and sketched in Fig. 1c.

The total current flowing in the metal phase is obtained by integrating $j_M$ from Eq. \eqref{jMjI} over $z$ from $0$ to $d(r)$:
\begin{equation}\label{IM}
I_M(r)=\frac{d(r)}{d(r)+(h-d(r))\rho_M/\rho_I}\frac{I}{\pi r}.
\end{equation}

In addition to Joule heating, we have Peltier heating; in fact $I_M$ changes with $r$, either because of the change in $d(r)$ or because of the spreading of the current, and thus some current flows across the phase interface.. The current density normal to the interface is given by $j_n=\partial_rI_M/\sqrt{1+(\partial_rd)^2}$, and produces an additional contribution $T(d)\Delta Sj_n(h-z)/\kappa$ to the temperature at the interface in Eq. \eqref{TsolUnif}, giving
\begin{equation}\label{TdPH}
T(d)=T_0+\Delta T\left(\frac{I}{j_{cr}\pi hr}\right)^2\left[1-\frac{d^2}{(d+(h-d)\rho_M/\rho_I)^2}\right]+T(d)\Delta S\frac{h-d}{\kappa}\frac{\partial_rI_M(r)}{\sqrt{1+(\partial_rd)^2}}.
\end{equation}

Equation \eqref{TdPH} is a differential equation for $d(r)$ and include both Peltier and Joule heating. Neglecting Joule heating corresponds to neglecting the first term in the square brackets. The derivative of $I_M(r)$ is
\begin{equation}\label{dIMr}
\partial_rI_M(r)=\frac{I}{\pi r}\left[\frac{h\rho_M/\rho_I}{(d(r)+(h-d(r))\rho_M/\rho_I)^2}\partial_rd-\frac{d(r)}{d(r)+(h-d(r))\rho_M/\rho_I}\frac1r\right].
\end{equation}

The second term is negligible in Eq. \eqref{dIMr} when $\partial_rd\gg d/r$; this is true near $r_c$ where $d\rightarrow0$ but is an approximation when $r$ gets smaller. We also have to consider that for $r\lesssim h$, the current has a significant $z$ component and the derivation that led to Eq. \eqref{dIMr} is not entirely correct anymore; nonetheless we neglect the second term for simplicity, write the interface condition $T(d)=T_{\rm{MIT}}$ and derive Eq. \eqref{TPel}
\begin{equation}\label{TdPH2}
T_{\rm{MIT}}=T_0+T_{\rm{MIT}}\Delta S\frac{I}{\pi r}\frac{\rho_M/\rho_I h^2}{\kappa(d+\rho_M/\rho_Ih)^2}\frac{\partial_rd}{\sqrt{1+(\partial_rd)^2}}.
\end{equation}

Integration of Eq. \eqref{TdPH2} for $I<0$ (and $\partial_rd\ll1$) leads to
\begin{equation}\label{rd}
\frac{\pi\kappa\Delta T}{2T_{\rm{MIT}}\Delta S|I|h^2}(r_c^2-r^2)\approx\frac{\rho_M/\rho_I}{d+(h-d)\rho_M/\rho_I}-\frac{\rho_M/\rho_I}{h\rho_M/\rho_I}=
\frac{d}{h(d+(h-d)\rho_M/\rho_I)}.
\end{equation}

For small distances $r\rightarrow0$, we assume that most of the current flows in the metal phase, which is equivalent to say that the resistance of the metal is smaller than that of the insulator, i.e. $d>(h-d)\rho_M/\rho_I$, so that we approximate the right-hand-side in Eq. \eqref{rd} and write
\begin{equation}\label{rc2}
\frac{\pi\kappa\Delta T}{2T_{\rm{MIT}}\Delta S|I|h^2}r_c^2\approx\frac1h\,\,\Rightarrow\,\,r_c=\sqrt{\frac{2T_{\rm{MIT}}\Delta S|I|h}{\pi\kappa\Delta T}},
\end{equation}
as written Eq. \eqref{rc}.

\end{document}